\documentclass[twocolumn,prb]{revtex4}
\usepackage{graphicx}
\usepackage{dcolumn}
\usepackage{bm}
\usepackage{amsmath}

\begin{document}

\preprint{}
\title{Theory of Josephson effect in chiral $p$-wave superconductor /
diffusive normal metal / chiral $p$-wave superconductor junctions }
\author{Y. Sawa$^1$, T. Yokoyama$^1$, Y. Tanaka$^1$, A. A. Golubov$^2$}
\affiliation{$^1$Department of Applied Physics, Nagoya University, Nagoya, 464-8603, Japan%
\\
and CREST, Japan Science and Technology Corporation (JST) Nagoya, 464-8603,
Japan \\
$^2$ Faculty of Science and Technology, University of Twente, 7500 AE,
Enschede, The Netherlands}
\date{\today}

\begin{abstract}
We study the Josephson effect between chiral $p$-wave superconductor
/ diffusive normal metal (DN) / chiral $p$-wave superconductor
(CP/DN/CP) junctions using quasiclassical Green's function formalism
with proper boundary conditions. The $p_{x}+ip_{y}$-wave symmetry of
superconducting order parameter is chosen which is believed to be a
pairing state in Sr$_2$RuO$_4$.
%The specific features of this system is that
%the mid gap Andreev resonant state (MARS) and the
%proximity effect coexist each other.
%
It is shown that the Cooper pairs induced in DN have an
odd-frequency spin-triplet $s$-wave symmetry, where pair amplitude
is an odd function of Matsubara frequency. Despite the peculiar
symmetry properties of the Cooper pairs, the behavior of the
Josephson current is rather conventional. We have found that the
current phase relation is almost sinusoidal and the Josephson
current is proportional to $\exp(-L/\xi)$, where $\xi$ is the
coherence length of the Cooper pair in DN and $L$ is the length of
DN. The Josephson current between CP / diffusive ferromagnet metal
(DF) / CP junctions is also calculated. It is shown that the 0-$\pi$
transition can be realized by varying temperature or junction length
$L$ similar to the case of conventional $s$-wave junctions. These
results may serve as a guide to study superconducting state of
Sr$_2$RuO$_4$.
\end{abstract}

\maketitle

%\pacs{PACS numbers: 74.45 +c, 74.50 +r.}

%--- title ---

%--- author ---

%
%--- address ---

%
%--- date ---

%   Abstract
%-----------------------------------------------------------

%-----------------------------------------------------------


\section{Introduction}

Exploration of unconventional superconducting pairing is one of
central issues in the physics of superconductivity. Possible
realization of spin-triplet superconductivity in Sr$_{2}$RuO$_{4}$
is widely discussed at present \cite{Maeno}. A number of
experimental results is consistent with spin-triplet chiral $p$-wave
symmetry state in this material \cite{Ishida,Luke,Mackenzie}. It is
well known that the midgap Andreev resonant state (MARS) \cite
{Buchholtz,Hara,Hu,Tanaka:did2} is induced near interfaces in
unconventional superconducting junctions where pair potential
changes its sign across the Fermi surface. The MARS manifests itself
as a zero bias conductance peak (ZBCP) in quasiparticle tunneling
experiments. A number of tunneling data in Sr$_{2}$RuO$_{4}$
junctions show ZBCP \cite{tunneling} in accordance with theoretical
predictions \cite{Yamashiro}. The Josephson
effect in Sr$_{2}$RuO$_{4}$ was also studied both theoretically \cite%
{Josephson} and experimentally \cite{jin}. Recent SQUID experiment by Nelson
\cite{Nelson} is consistent with the realization of chiral $p$-wave
superconducting state \cite{AsanoSQUID}.

At the same time, there are important recent achievements in
theoretical study of the proximity effect in  junctions between
unconventional superconductors. 
%
%There are a lot of interesting phenomena predicted
%in triplet superconductor (TS) / diffusive normal metal (DN) hybrid junctions.
%In these junctions, mid gap Andreev resonant state
%(MARS)\cite{Buchholtz,Hara,Hu,Tanaka:did2}
%is induced near the interface and penetrates into the DN region.
It was predicted that in diffusive normal metal (DN) / triplet
superconductor (TS) junctions MARS formed at the DN/TS interface,
can penetrate into DN \cite{p-wave}.
%The proximity effect which is the Cooper-pairs penetrate from
%the superconductor, and MARS specific in TS coexist in these junctions.
This proximity effect is very unusual since it generates the zero
energy peak (ZEP) in the local density of states (LDOS) in contrast
to the conventional proximity effect where the resulting LDOS has a
minigap \cite {Proximity}. It was also shown theoretically, that the
ZEP appears in the chiral $p$-wave state. Thus to explore the ZEP in
DN region of DN/Sr$_{2}$RuO$_{4}$ heterostructures is an intriguing
topic \cite{p-wave}. Very recently, it is predicted that the induced
Cooper pairs in DN are in an unconventional odd-frequency symmetry
state, in contrast to the usual even-frequency pairing
\cite{Odd-Golubov}. Since this proximity effect specific to TS is
completely new phenomenon, it is very interesting to study the
Josephson effect in TS/DN/TS junctions.

Recently, it is shown that the Josephson current is enhanced
strongly at low temperatures in TS/DN/TS junctions\cite{Asano:pdnp}
and is proportional to sin$(\Psi/2)$, where $\Psi$ is a
superconducting phase difference between left and right
superconductors \cite{Asano2006}. These results are quite different
from those for $d$-wave superconductor / DN / $d$-wave superconductor
junctions \cite{Asanod-wave}. However, in most of previous theories
of TS/DN/TS junctions, only the $p$-wave state in the presence of
the time reversal symmetry was considered. The existing knowledge of
the Josephson effect in TS/DN/TS junctions for chiral $p$-wave
symmetry is very limited~\cite{asano02-2}. It is important to study
the Josephson effect in the chiral $p$-wave junctions in much more
detail because this symmetry is the most promising superconducting
state in Sr$_{2}$RuO$_{4}$.

To study this problem, the quasi-classical Green's function theory
is the useful method. In diffusive regime, the quasi-classical
Green's function obeys the Usadel equations\cite{Usadel}. The
circuit theory\cite {Nazarov1} enables one to treat the case of
arbitrary interface transparency in $s$-wave superconductor (S)
junctions. This theory was recently generalized for unconventional
superconductor (US) junctions\cite
{PRL2003,Tanaka2004,Tanaka2005,Yokoyama1}. In these approach, the
effect of MARS is naturally included. The theory was extended to the
cases of US/DN/US and US/diffusive ferromagnet (DF)/US junctions
where time reversal symmetry is present in US \cite{Yokoyama,Yoko4}.
However, these theories can not be applied to calculating the
Josephson current in chiral $p$-wave superconductor / DN / chiral
$p$-wave superconductor (CP/DN/CP) junctions with
$p_{x}+ip_{y}$-wave symmetry of the pair wave function in chiral
$p$-wave superconductor. The purpose of this paper is to generalize
the above approach and apply it to the interface between DN (DF) /
superconductor with broken time reversal symmetry.

In the present paper, we derive the boundary conditions of
quasiclassical Green's function available for DN (DF) / CP interface
in the presence of the Josephson effect and calculate the Josephson
current in CP / DN (DF) / CP junctions by solving the Usadel
equations with these boundary conditions. It is shown that the
induced pair in DN is purely in the odd-frequency pairing state. The
magnitude of the calculated Josephson current is larger than that in the
$s$-wave superconductor / DN / $s$-wave superconductor (S/DN/S) junctions.
However, it is smaller than
that in $p_{x}$-wave superconductor / DN / $p_x$-wave superconductor (P/DN/P)
junctions. %
%Then the magnitude of the induced pair amplitude in DN is
%not enhanced as compared to $p_{x}$-wave superconductor
%junctions.
%
The obtained temperature dependence of the Josephson current is
similar to that in the conventional $s$-wave junctions. The current
phase relation is almost sinusoidal and the Josephson current is
proportional to $\exp(-L/\xi)$, where $\xi$ is the coherence length
of the Cooper pair in DN and $L$ is the width of DN. We have also
calculated the Josephson current in CP/DF/CP junctions. Similar to
the case of the S/DF/S junctions, the $0-\pi$ transition occurs as a function of
the length of DF. As follows from these results, it is difficult to
extract the unusual properties of proximity effect specific to
spin-triplet $p$-wave superconductor junctions if we look at d.c.
Josephson effect only. These results may serve as a guide to explore
novel properties in superconducting Sr$_{2}$RuO$_{4}$.

\section{Formulation}

\begin{figure}[htbp]
\begin{center}
\includegraphics[keepaspectratio=true,height=45mm]{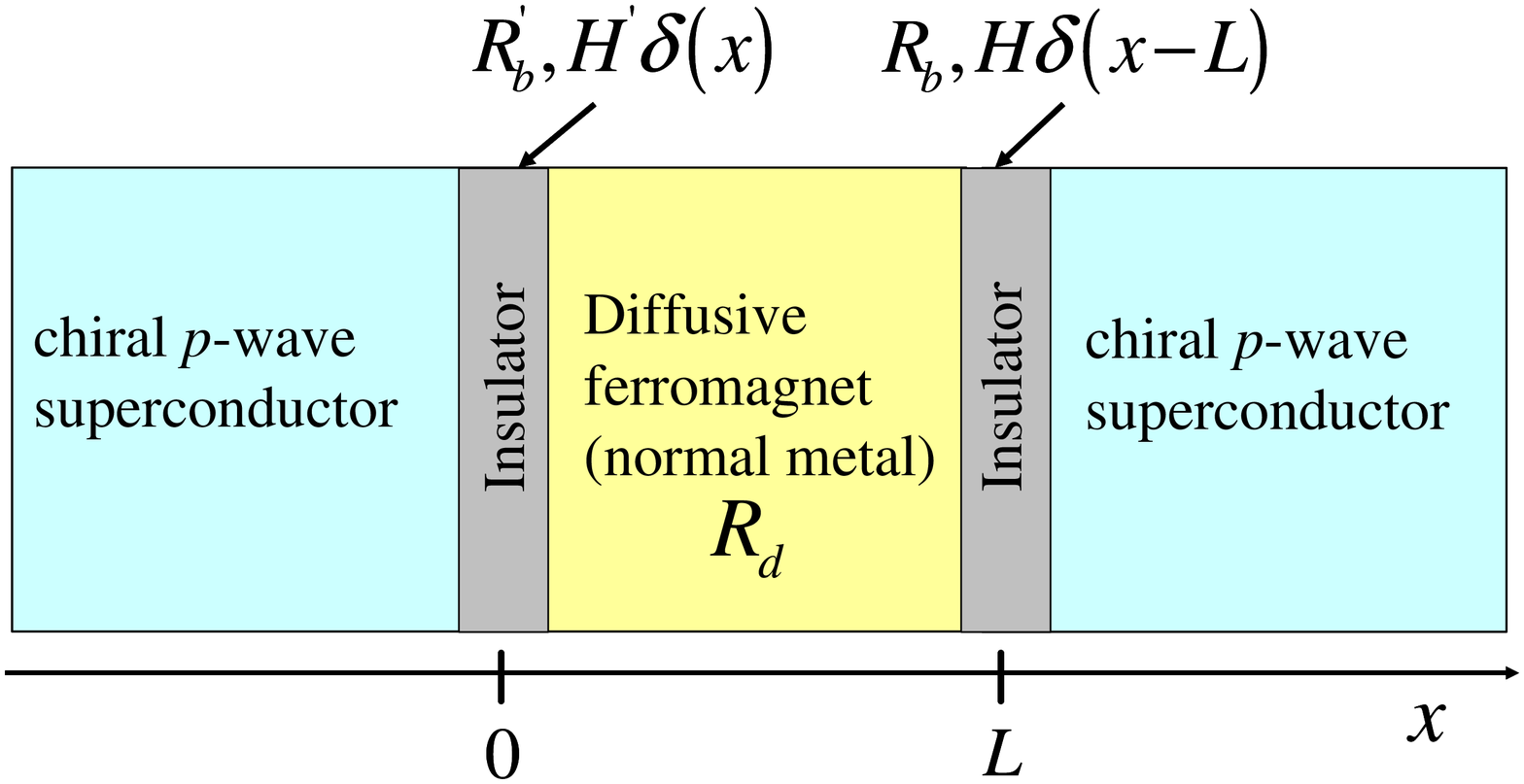}
\end{center}
\caption{(Color online) Schematic illustration of the model.}
\label{chiral}
\end{figure}

In the following sections, the units with $\hbar=k_{B}=1$ are used.
The model of CP / DF (DN) / CP junction is illustrated in
Fig.\ref{chiral}. Here $ R^{^{\prime}}_{b}$ is a resistance of
insulating barrier located at $x=0$, $ R_d$ is a resistance of the
DN, $R_b$ is a resistance of insulating barrier located at $x=L$,
and the length of DN $L$ is much larger than the mean flee path. The
infinitely narrow insulating barriers are modeled as $
U(x)=H^{\prime}\delta(x)+H\delta(x-L)$. Then the barrier
transparency $T^{(^{\prime})}_{m}$ is expressed by
$T^{(^{\prime})}_{m}=4\cos^2\phi/(4\cos^{2}\phi
+Z^{(^{\prime})2})$ with $Z^{(^{\prime})}=2H^{(^{\prime})}/ v_F$%
. Here $\phi$ is injection angle measured from the direction perpendicular
to the interface between DF (DN) and chiral superconductor, and $v_F$ is
Fermi velocity.

First, we concentrate on the Nambu-Keldysh (NK) Green's function in DF (DN)
within the quasiclassical approximation. We define NK Green's function as $%
\check{G}_{N}(x)$. We denote the retarded part of $\check{G}_{N}(x)$ as 
$\hat{R}_{N}(x)$, which is given by
\begin{equation}
\hat{R}_{N}=\sin\theta \cos\psi \hat{\tau}_{1} +\sin\theta \sin\psi \hat{\tau%
}_{2} +\cos\theta \hat{\tau}_{3},
\end{equation}
where $\hat{\tau}_{j}$ (j=1,2,3) are the Pauli matrices in electron-hole
space.

The functions $\theta$ and $\psi$ for majority (minority) spins obey
the Usadel equation:
\begin{eqnarray}
D\left[\frac{\partial^{2}}{\partial x^{2}}\theta - \left(\frac{\partial \psi%
}{\partial x}\right)^{2}\cos \theta \sin \theta\right] + 2i(\varepsilon
+(-)h) \sin\theta =0,  \notag
\end{eqnarray}
\begin{eqnarray}
\frac{\partial}{\partial x} \left[\sin^{2}\theta \left(\frac{\partial \psi}{%
\partial x}\right)\right]=0,
\end{eqnarray}
with the diffusion constant $D$ and the exchange field $h$. If we
choose $h =0 $, DF is reduced to be DN. The boundary condition for
$\check{G}_{N}(x)$ at DF (DN) / CP interface has the form
\begin{eqnarray}
\frac{L}{R_{d}}\left.\left[\check{G}_{N}(x)\frac{\partial \check{G}_{N}(x)}{%
\partial x} \right]\right|_{x=L_{-}}=-\frac{\langle \check{I}_{m} \rangle}{%
R_{b}},  \notag
\end{eqnarray}
\begin{equation*}
\check{I}(\phi)=\check{I}_{m}=2\left[\check{G}_{1},\check{B}_{m}%
\right],
\end{equation*}
\begin{eqnarray}
\check{B}_{m}=\left(-T_{1m}[\check{G}_{1},\check{H}_{-}^{-1}]+\check{H}%
_{-}^{-1} \check{H}_{+}-T_{1m}^{2}\check{G}_{1}\check{H}_{-}^{-1}\check{H}%
_{+} \check{G}_{1}\right)^{-1}  \notag
\end{eqnarray}
\begin{eqnarray}
\times \left(T_{1m}(1-\check{H}_{-}^{-1})+T_{1m}^{2}\check{G}_{1} \check{H}%
_{-}^{-1}\check{H}_{+}\right),  \label{Nazarov}
\end{eqnarray}
with $\check{G}_{1}$=$\check{G}_{N}(x=L_{-})$, $\check{H}_{\pm}$=$(\check{G}%
_{2+} \pm \check{G}_{2-})/2$, and
$T_{1m}=T_{m}/(2-T_{m}+2\sqrt{1-T_{m}})$. Here $\check{G}_{2\pm}$ is
the asymptotic Green's function in the superconductor defined as in
the previous paper\cite{PRL2003}.

Here, the average over the various angles of injected particle at the
interface is defined as
\begin{eqnarray}
\langle \check{I}_{m} \rangle= \frac{ \int_{-\pi /2}^{\pi /2}d\phi\; \cos
\phi \;\check{I}_{m}}{ \int_{-\pi /2}^{\pi /2}d\phi\; \cos \phi\; T_m }.
\end{eqnarray}

Then the resistance of the interface $R^{(^{\prime})}_b$ is given by
\begin{eqnarray}
R^{(^{\prime})}_{b}=\frac{2R^{\prime}_{0}}{\int_{-\pi /2}^{\pi /2}d\phi\;
\cos\phi\; T^{(^{\prime})}_m },
\end{eqnarray}
where $R^{(^{\prime})}_0$ is the Sharvin resistance, which in the three
dimensional case is expressed by $R^{(^{\prime})-1}_{0}=e^{2}k_{F}^{2}S^{^{%
\prime}}_{c}/4\pi^2$. Here, $k_{F}$ is the Fermi wave number and $%
S^{^{\prime}}_{c}$ is the constriction area.

Retarded component of $\check{G}_{2\pm}$, $i.e.$, $\hat{R}_{2\pm}$, is
expressed by
\begin{eqnarray}
\hat{R}_{2\pm}=(f_{1\pm}\cos\Psi+f_{2\pm}\sin\Psi)\hat{\tau}_{1}  \notag \\
+(f_{1\pm}\sin\Psi-f_{2\pm}\cos\Psi) \hat{\tau}_{2} + g_\pm \hat{\tau}_{3},
\end{eqnarray}
with $f_{1\pm}$=Re$(f_\pm)$, $f_{2\pm}$=Im$(f_\pm)$, $g_{\pm}=\epsilon/\sqrt{%
\epsilon^2-|\Delta_{\pm}|^{2}}$, $f_{\pm}=\Delta_{\pm}/\sqrt{%
|\Delta_{\pm}|^{2}-\epsilon^2}$, and the macroscopic phase of the
superconductor $\Psi$. Here, $\Delta_{+}=\Delta(\phi)$ and $%
\Delta_{-}=\Delta(\pi-\phi)$ are the pair potentials corresponding
to the injection angles $\phi$ and $\pi-\phi$ respectively. Note
that $|\Delta_{+}|$=$ |\Delta_{-}|$ is satisfied in the present
case, then we can put $ g_{+}=g_{-}\equiv g$.

Next we consider the boundary condition for the retarded part of the
NK Green's functions at the DF (DN) / CP interface.
The left side of the boundary condition of Eq.(\ref{Nazarov}) is
expressed by

\begin{eqnarray}
\frac{L}{R_{d}}\hat{R}_{N}(x)\frac{\partial}{\partial x}
\hat{R}_{N}(x) |_{x=L_{-}}\nonumber
\end{eqnarray}
\begin{eqnarray}
=\frac{Li}{R_d}\bigg{[}\bigg{(}-\frac{\partial \theta}{\partial
x}\sin \psi-\frac{\partial \psi}{\partial x} \sin \theta \cos \theta
\cos \psi \bigg{)} \hat{\tau}_{1}\nonumber
\end{eqnarray}
\begin{eqnarray}
+ \bigg{(}\frac{\partial \theta}{\partial x}\cos \psi-\frac{\partial
\psi}{\partial x} \sin \theta \cos \theta \sin \psi \bigg{)}
\hat{\tau}_{2} +\frac{\partial \psi}{\partial x} \sin^{2}\theta
\hat{\tau}_{3}\bigg{]}.
\end{eqnarray}

In the right side of Eq.(\ref{Nazarov}), $\hat{I}_R$ is expressed by

\begin{eqnarray}
\hat{I}_{R}=4iT_{1m}(\mathbf{d}_{R}  \cdot \mathbf{d}_{R})^{-1}
\nonumber
\end{eqnarray}
\begin{eqnarray}
\times\biggl(-\frac{1}{2}(1 + T_{1m}^{2})({\mathbf s}_{2+}-{\mathbf
s}_{2-})^{2} [{\mathbf s}_{1}\times({\mathbf s}_{2+} + {\mathbf
s}_{2-})]\cdot \hat{\mathbf{\tau}} \nonumber
\end{eqnarray}
\begin{eqnarray}
+2T_{1m}{\mathbf s}_{1}\cdot({\mathbf s}_{2+} \times {\mathbf
s}_{2-}) [{\mathbf s}_{1} \times ({\mathbf s}_{2+} \times {\mathbf
s}_{2-})] \cdot \hat{\mathbf{\tau}} \nonumber
\end{eqnarray}
\begin{eqnarray}
+2T_{1m}{\mathbf s}_{1}\cdot({\mathbf s}_{2+} - {\mathbf s}_{2-})
[{\mathbf s}_{1} \times ({\mathbf s}_{2+} - {\mathbf s}_{2-})] \cdot
\hat{\bf \tau}\nonumber
\end{eqnarray}
\begin{eqnarray}
-i(1+T_{1m}^{2})(1 -{\mathbf s}_{2+}\cdot{\mathbf s}_{2-}) [{\mathbf
s}_{1} \times ({\mathbf s}_{2+} \times {\mathbf s}_{2-})]\cdot
\hat{\mathbf{\tau}}\nonumber
\end{eqnarray}
\begin{eqnarray}
+2iT_{1m}(1 -{\mathbf s}_{2+}\cdot{\mathbf s}_{2-}) [{\mathbf
s}_{1}\cdot({\mathbf s}_{2+}-{\mathbf s}_{2-}){\mathbf s}_{1}
\nonumber
\end{eqnarray}
\begin{eqnarray}
-({\mathbf s}_{2+} -{\mathbf s}_{2-})]\cdot
\hat{\mathbf{\tau}}\biggl),
\end{eqnarray}
\begin{eqnarray}
\mathbf{d}_{R} =(1 + T_{1m}^{2})(\mathbf{s}_{2+} \times
\mathbf{s}_{2-})
-2T_{1m}\mathbf{s}_{1}\times(\mathbf{s}_{2+}-\mathbf{s}_{2-})\nonumber\\
-2T_{1m}^{2}\mathbf{s}_{1}\cdot(\mathbf{s}_{2+}\times\mathbf{s}_{2-})\mathbf{s}_{1},
\end{eqnarray}
with $\hat{R}_{1}={\mathbf s}_{1}\cdot\hat{\mathbf{\tau}}$, and
$\hat{R}_{2\pm}={\mathbf s}_{2\pm}\cdot\hat{\mathbf{\tau}}$. Here,
$\check{I}_{R}$ is the retarded part of $\check{I}_{m}$. The
spectral vector ${\mathbf s}_{1}$, and ${\mathbf s}_{2\pm}$ are
expressed by
\begin{eqnarray}
{\mathbf s}_{1} = \left(
\begin{array}{c}
\sin\theta\cos\psi  \\
\sin\theta\sin\psi  \\
\cos \theta
\end{array}
\right),\nonumber
\end{eqnarray}
\begin{eqnarray}
\mathbf{s}_{2\pm}= \left(
\begin{array}{c}
f_{1\pm}\cos\Psi+f_{2\pm}\sin\Psi  \\
f_{1\pm}\sin\Psi-f_{2\pm}\cos\Psi  \\
g
\end{array}
\right).
\end{eqnarray}
Here, $\Psi$ is the macroscopic phase of right superconductor. After
some algebra,  the matrix current is reduced to be
\begin{eqnarray}
\hat{I}_{R} =2iT_{m}[(2-T_{m})+T_{m}(g_{s}\cos\theta +
f_{s}\sin\theta \sin(\psi-\Psi))]^{-1}\nonumber
\end{eqnarray}
\begin{eqnarray}
\times([-g_{s} \sin\theta \sin\psi-f_{s}\cos\theta \cos\Psi]
\hat{\tau}_{1}\nonumber
\end{eqnarray}
\begin{eqnarray}
+[g_{s} \sin\theta \cos\psi-f_{s}\cos\theta \sin\Psi]
\hat{\tau}_{2}\nonumber
\end{eqnarray}
\begin{eqnarray}
+f_{s}\sin\theta \cos(\psi-\Psi) \hat{\tau}_{3})
\end{eqnarray}

Then the resulting $\hat{I}_{R}$ can be expressed as
\begin{eqnarray}
\hat{I}_{R} = \left(
\begin{array}{c}
-I_{1} \sin\theta \sin\psi-I_{2}\cos\theta \cos\Psi   \\
I_{1} \sin\theta \cos\psi-I_{2}\cos\theta \sin\Psi  \\
I_{2}\sin\theta \cos(\psi-\Psi)
\end{array}
\right),
\end{eqnarray}

\begin{eqnarray}
I_1=\left<\frac{2T_m g_s}{\Lambda_m}\right>, I_2=\left<\frac{2T_m
f_s}{\Lambda_m}\right>,\nonumber
\end{eqnarray}
\begin{eqnarray}
\Lambda_{m}=2-T_m +T_m \left[g_{s}\cos\theta-f_{s}\sin\theta
\sin(\psi-\Psi)\right], \nonumber
\end{eqnarray}
\begin{eqnarray}
g_s=\frac{2g+i(f_{1+}f_{2-}-f_{2+}f_{1-})}{1+g^2+f_{1+}f_{1-}+f_{2+}f_{2-}},\nonumber\\
f_s=\frac{ig(f_{1+}-f_{1-})+f_{2+}+f_{2-}}{1+g^2+f_{1+}f_{1-}+f_{2+}f_{2-}}.
\label{Boundary}
\end{eqnarray}

Finally the boundary condition is obtained as
\begin{eqnarray}
\frac{LR_{b}}{R_{d}} \frac{\partial \theta}{\partial x}
=-I_{1}\sin\theta-I_{2}\cos\theta \sin(\psi-\Psi), \nonumber
\end{eqnarray}

\begin{eqnarray}
\frac{LR_{b}}{R_{d}} \frac{\partial \psi}{\partial x} \sin\theta
=-I_{2}\cos(\psi-\Psi).
\end{eqnarray}

%Similarly, boundary condition at singlet chiral superconductor / DF (DN)
%interface is obtained by putting $\Psi$ to $\Psi+\pi/2$,
%and
%\begin{eqnarray}
%f_s=\frac{f_{1+}+f_{1-}+ig(f_{2+}-f_{2-})}{1+g^2+f_{1+}f_{1-}+f_{2+}f_{2-}}.
%\end{eqnarray}

%These boundary conditions are very general
%because they are applicable
%for any superconductor / DF(DN) interface with $g_{+}=g_{-}=g$,
%including both with and without
%broken  time reversal symmetry cases.
%
For the calculation of thermodynamically quantities, it is
convenient to use Matsubara representation by changing $\epsilon$
$\to$ $i\omega$.

We parameterize the quasiclassical Green's functions $G_{\omega}$
and $F_{\omega}$ with a function $\Phi_{\omega}$,
\cite{Likharev,Golubov}
\begin{equation}
G_{\omega}=\frac{\omega}{\sqrt{\omega^2+\Phi_{\omega}\Phi^*_{-\omega}}},\
\
F_{\omega}=\frac{\Phi_{\omega}}{\sqrt{\omega^2+\Phi_{\omega}\Phi^{*}_{-\omega}}},
\end{equation}
where $\omega$ is Matsubara frequency. The following relations are
satisfied:
\begin{eqnarray}
\frac{G_{\omega}}{2\omega}\left(\Phi_{\omega}+\Phi^*_{-\omega}\right) = \sin\theta\cos\psi,\nonumber\\
\frac{iG_{\omega}}{2\omega}\left(\Phi_{\omega}-\Phi^*_{-\omega}\right)
= \sin\theta\sin\psi.
\end{eqnarray}

Then the Usadel equation for majority (minority) spin has the
form\cite{Golubov}
\begin{equation}
\xi^2\frac{\pi T_{\rm{C}}}{G_{\omega}}\frac{\partial }{\partial
x}\left(G^2_{\omega}\frac{\partial }{\partial
x}\Phi_{\omega}\right)-(\omega -(+)ih) \Phi_{\omega}=0,
\end{equation}
with the coherence length $\xi=\sqrt {D/2\pi T_{\rm{C}}}$, the diffusion
constant $D$, the exchange field $h$, and the transition temperature
$T_{\rm{C}}$.

The boundary condition at $x=L$ is expressed by
\begin{eqnarray}
\frac{G_{\omega}}{\omega}\frac{\partial }{\partial
x}\Phi_{\omega}=\frac{R_d}{R_bL}\left(-\frac{\Phi_{\omega}}{\omega}I_1+ie^{-i\Psi}I_2
\right), \nonumber \end{eqnarray}
\begin{eqnarray}
I_1=\left<\frac{2T_m g_s}{\Lambda_m}\right>, I_2=\left<\frac{2T_m
f_s}{\Lambda_m}\right>,\nonumber
\end{eqnarray}
\begin{eqnarray}
\Lambda_{m}=2-T_m +T_m
\left[g_sG_\omega+f_s\left(B\sin\Psi-C\cos\Psi\right)\right],
\nonumber
\end{eqnarray}
\begin{eqnarray}
B=\frac{G_\omega}{2\omega}\left(\Phi_{\omega}+\Phi^*_{-\omega}\right),
C=\frac{iG_\omega}{2\omega}\left(\Phi_{\omega}-\Phi^*_{-\omega}\right).
\end{eqnarray}

The boundary condition at $x=0$ is expressed by
\begin{eqnarray}
\frac{G_{\omega}}{\omega}\frac{\partial }{\partial
x}\Phi_{\omega}=-\frac{R_d}{R^{'}_bL}\left(-\frac{\Phi_{\omega}}{\omega}I'_1+iI'_2
\right).
\end{eqnarray}
Here  $I'_1$ and $I'_2$ are obtained by adding subscript ', changing
$\phi$ to $\pi - \phi$, and putting $\Psi=0$ for $I_1$ and $I_2$ at
$x=L$. Then the macroscopic phase differences between 
left and right superconductor becomes $\Psi$.\par
To discuss the features of the proximity effect, in the following we
will study the frequency dependence of the induced pair amplitude in
DN by choosing  $h=0$.
Before proceeding with formal discussion, let us present qualitative
arguments. Two constrains should be satisfied in the considered
system: (1) only the $s$-wave even-parity state is possible in the
DN due to isotropization by impurity scattering, (2) the spin
structure of induced Cooper pairs in the DN is the same as in an
attached superconductor. Then the Pauli principle provides the
unique relations between the pairing symmetry in a superconductor
and the resulting symmetry of the induced pairing state in the DN
\cite{Odd-Golubov}.
Since there is no spin flip at the interface, it is natural to
expect that the odd-frequency pairing state is generated in DN.
It is possible to show that
\begin{equation}
f_{\pm}(-\omega)=f_{\pm}(\omega), \
g_{\pm}(-\omega)=-g_{\pm}(\omega)
\end{equation}
Using these equations,
\begin{eqnarray}
g_{s}(-\omega,-\phi)=-g_{s}(\omega,\phi),\
f_{s}(-\omega,-\phi)=-f_{s}(\omega,\phi).
\end{eqnarray}
are satisfied. For $h=0$, the Usadel equation has the
form\cite{Golubov}
\begin{equation}
\xi^2\frac{\pi T_{\rm{C}}}{G_{\omega}}\frac{\partial }{\partial
x}\left(G^2_{\omega}\frac{\partial }{\partial
x}\Phi_{\omega}\right)-\omega  \Phi_{\omega}=0,
\end{equation}
and the boundary condition at $x=L$ is
\begin{eqnarray}
\frac{G_{\omega}}{\omega}\frac{\partial }{\partial x}
\Phi_{\omega}=\frac{R_d}{R_bL}
\left(-\frac{\Phi_{\omega}}{\omega}I_1(\omega,\phi)
+ie^{-i\Psi}I_2(\omega,\phi) \right), \nonumber
\end{eqnarray}
\begin{eqnarray}
I_1=\left<\frac{2T_m g_s}{\Lambda_m}\right>, I_2=\left<\frac{2T_m
f_s}{\Lambda_m}\right>,\nonumber
\end{eqnarray}
\begin{eqnarray}
\Lambda_{m}=2-T_m +T_m
\left[g_sG_\omega+f_s\left(B\sin\Psi-C\cos\Psi\right)\right],
\nonumber
\end{eqnarray}
\begin{eqnarray}
B=\frac{G_\omega}{2\omega}\left(\Phi_{\omega}+\Phi^*_{-\omega}\right),
C=\frac{iG_\omega}{2\omega}\left(\Phi_{\omega}-\Phi^*_{-\omega}\right).
\end{eqnarray}
By changing $\omega$ and $\phi$ into $-\omega$ and $-\phi$ in eqs.
(23) and (24), following equations are obtained
\begin{equation}
\xi^2\frac{\pi T_{\rm{C}}}{G_{-\omega}}\frac{\partial }{\partial
x}\left(G^2_{-\omega}\frac{\partial }{\partial
x}\Phi_{-\omega}\right)+\omega \Phi_{-\omega}=0,
\end{equation}

\begin{eqnarray}
\frac{G_{-\omega}}{\omega}\frac{\partial }{\partial
x}\Phi_{-\omega}=\frac{R_d}{R_bL}
\left[\frac{\Phi_{-\omega}}{\omega}I_1(-\omega,-\phi)
+ie^{-i\Psi}I_2(-\omega,-\phi) \right]. \nonumber
\end{eqnarray}
To check the consistency of the four above equations, we consider
the $\omega$ dependence of several quantities. One can show that
\begin{eqnarray}
f_{1\pm}(-\omega)=f_{1\pm}(\omega), \
f_{2\pm}(-\omega)=f_{2\pm}(\omega), \nonumber \\
g(-\omega)=-g(\omega)
\end{eqnarray}
As a result,
\begin{equation}
g_{s}(-\omega,-\phi)=-g_{s}(\omega,\phi) \
f_{s}(-\omega,-\phi)=-f_{s}(\omega,\phi)
\end{equation}
By comparing Eqs. (23) with Eq. (25), we can derive
\begin{equation}
G_{-\omega}=-G_{\omega}
\end{equation}
Two cases can be considered: \par (1)
\begin{eqnarray}
\Phi_{-\omega}=\Phi_{\omega}, \
I_{1}(-\omega,-\phi)=-I_{1}(\omega,\phi) \nonumber \\
I_{2}(-\omega,-\phi)=I_{2}(\omega,\phi)
\end{eqnarray}
(2)
\begin{eqnarray}
\Phi_{-\omega}=-\Phi_{\omega}, \
I_{1}(-\omega,-\phi)=-I_{1}(\omega,\phi) \nonumber \\
I_{2}(-\omega,-\phi)=-I_{2}(\omega,\phi)
\end{eqnarray}
For the case (1), the relations $B(-\omega)=B(\omega)$ and
$C(-\omega)=C(\omega)$ hold, while for case (2) the relations
$B(-\omega)=-B(\omega)$ and $C(-\omega)=-C(\omega)$ hold.
For the case (1), since $\Lambda_{m}(-\omega,-\phi) \neq
\Lambda_{m}(\omega,\phi)$, then it is impossible to satisfy
$I_{1}(-\omega,-\phi)=-I_{1}(\omega,\phi)$ and
$I_{2}(-\omega,-\phi)=I_{2}(\omega,\phi)$ simultaneously, thus, this
case can not be realized. For the case (2),
\begin{equation}
\Lambda_{m}(-\omega,-\phi)=\Lambda_{m}(\omega,\phi),
\end{equation}
is satisfied and this  relation is consistent with
$I_{1}(-\omega,-\phi)=-I_{1}(\omega,\phi)$
$I_{2}(-\omega,-\phi)=-I_{2}(\omega,\phi)$. Since
$\Phi(\omega)=-\Phi(\omega)$ is satisfied, we can show
\begin{eqnarray}
\sin \theta(-\omega)\cos \psi(-\omega)=
-\sin \theta(\omega)\cos \psi(\omega),\nonumber\\
\sin \theta(-\omega)\sin \psi(-\omega)= -\sin \theta(\omega)\sin
\psi(\omega).
\end{eqnarray}
Then $F_{-\omega}=-F_{\omega}$ is satisfied. This indicates the
realization of the odd-frequency pairing state in DN. In the
presence of $h$, $i.e.$, CP/DF/CP junctions, the admixture of
even-frequency spin-singlet even-parity state is also present.
\par

The Josephson current is given by
\begin{equation}
\frac{eIR}{\pi T_{\rm{C}}}=i\frac{RTL}{4R_dT_{\rm{C}}}\sum_{\uparrow, \downarrow
,\omega}\frac{G^2_\omega}{\omega^2}\left( \Phi_\omega\frac{\partial
}{\partial x}\Phi^*_{-\omega}-\Phi^*_{-\omega}\frac{\partial
}{\partial x}\Phi_\omega\right),
\end{equation}
where $T$ is temperature, and $R=R_b+R'_b+R_d$ is the total
resistance of the junction. In the following, we fix $R'_b$=$R_b$,
$T'_m=T_m$, and choose $\Delta(\phi)=\Delta e^{i\phi}$. Here, we
define $\Delta_0$ as $\Delta_{0}\equiv \Delta(0)$.

\section{Results}

\subsection{CP / DN / CP junctions}

\begin{figure}[htbp]
  \begin{center}
\includegraphics[keepaspectratio=true,height=90mm]{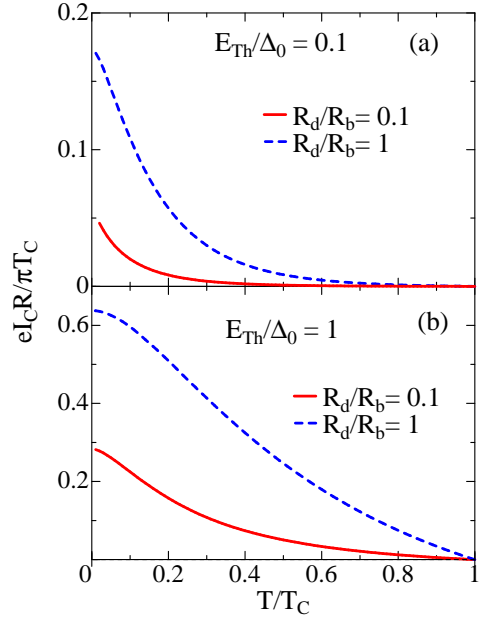}
  \end{center}
  \caption{(Color online) Temperature dependence of the maximum Josephson current for $Z=10$.
  The solid lines are the results for $R_{d}/R_{b}=0.1$ and broken lines are the results for $R_{d}/R_{b}=1$.
  (a) $E_{\rm{Th}}/\Delta_{0}=0.1$ and (b) $E_{\rm{Th}}/\Delta_{0}=1$.}
  \label{CC1}
\end{figure}

First, we consider the temperature dependence of a maximum Josephson
current $I_{\rm{C}}$ for $Z=10$ as shown in Fig.\ref{CC1}.
The magnitude of $I_{\rm{C}}$ is enhanced for large $E_{\rm{Th}}/\Delta_{0}$
and large $R_{d}/R_{b}$. It is  enhanced at low temperatures in both
cases (a) and (b). These features are consistent with conventional
case of S/ DN /S junctions.

\begin{figure}[htbp]
  \begin{center}
\includegraphics[keepaspectratio=true,height=90mm]{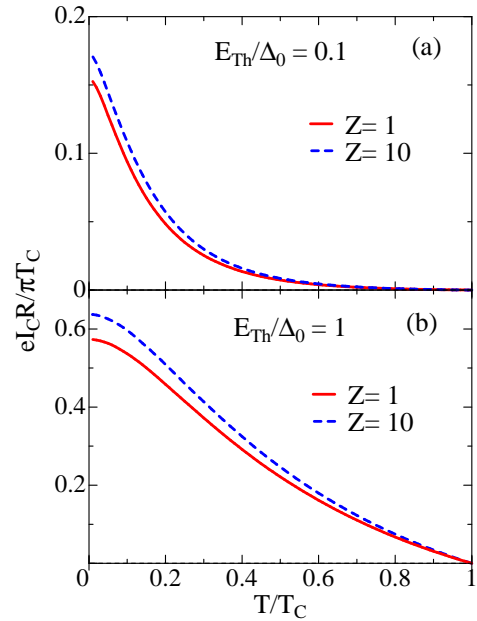}
  \end{center}
  \caption{(Color online) Temperature dependence of the maximum Josephson current for $R_{d}/R_{b}=1$.
  The solid lines are the results for $Z=1$ and the broken lines are the results for $Z=10$.
  (a) $E_{\rm{Th}}/\Delta_{0}=0.1$ and (b) $E_{\rm{Th}}/\Delta_{0}=1$.}
  \label{CC2}
\end{figure}

Next, we consider the dependence of $I_{\rm{C}}$ on $Z$, the transparency
parameter at the interface. Figure \ref{CC2} shows the temperature
dependence of the critical Josephson current for $R_{d}/R_{b}$= 1.
The magnitude of $RI_{\rm{C}}$ is enhanced for large $Z$, i.e., low
transparent interface for both $E_{\rm{Th}}/\Delta_{0}=0.1$ and 1. This
result is specific for junctions between triplet superconductors,
where proximity effect is enhanced by MARS formed at the interface.
It is known that the degree of the influence of MARS on the charge
transport becomes prominent for low transparent junctions with large
$Z$ \cite{p-wave}. On the contrary, in S/DN/S junctions the maximum
Josephson current is suppressed for large $Z$ \cite{Asano2006}.

\begin{figure}[htbp]
  \begin{center}
\includegraphics[keepaspectratio=true,height=90mm]{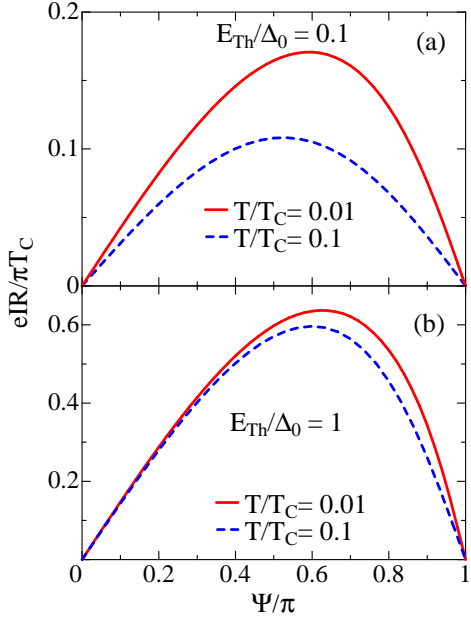}
  \end{center}
  \caption{(Color online) The current-phase relation for $Z=10$ and $R_d/R_b=1$.  The solid lines are the results for $T/T_{\rm{C}}=0.01$ and the broken lines are the results for $T/T_{\rm{C}}=0.1$.
  (a) $E_{\rm{Th}}/\Delta_{0}=0.1$ and (b)$E_{\rm{Th}}/\Delta_{0}=1$.}
  \label{CPR1}
\end{figure}

Next, we study the current-phase relation in order to examine the
unusual proximity effect specific to CP/DN/CP junctions. Figure
\ref{CPR1} shows the current-phase relation for $Z=10$ and $R_d/R_b=1$. 
We find that the peak is shifted to $\Psi>0.5\pi$ at low
temperatures, and this effect becomes rather strong in particular
for large $E_{\rm{Th}}/\Delta_{0}$. The result indicates that the
magnitude of the Josephson current is enhanced by the proximity
effect, and the Josephson current is not proportional to sin$\Psi$.
However this effect is not as pronounced as in the case of
$p_{x}$-wave /DN/$p_{x}$-wave (P/DN/P) junctions\cite{Asano2006}.

\begin{figure}[htbp]
  \begin{center}
\includegraphics[keepaspectratio=true,height=90mm]{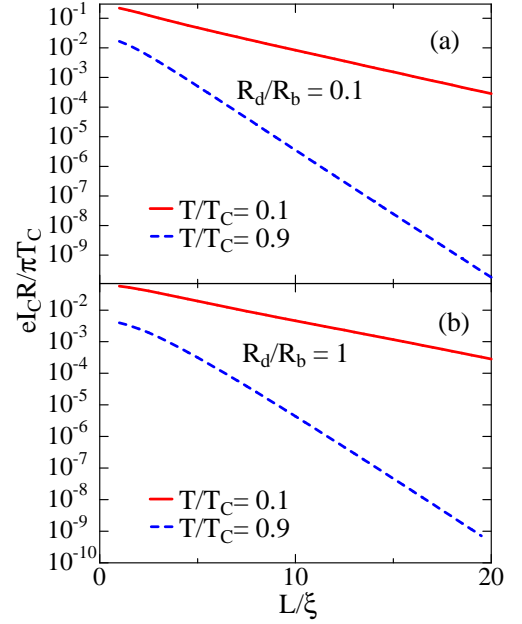}
  \end{center}
  \caption{(Color online) The critical current as a function of DN length for $Z=10$.
  The solid lines are the results for $T/T_{\rm{C}}=0.1$ and the broken lines are the results for $T/T_{\rm{C}}=0.9$.
  (a) $R_d/R_b=0.1$ and (b) $R_d/R_b=1$.}
  \label{Leng1}
\end{figure}

The dependence of the $I_{\rm{C}}$ on the length of DN is shown in Figure
\ref{Leng1} for $Z=10$ and $R_d/R_b=1$. We find that the $I_{\rm{C}}$
is proportional to $\exp(-L/\xi)$ in agreement with existing
theoretical results.

\begin{figure}[htbp]
  \begin{center}
\includegraphics[keepaspectratio=true,height=90mm]{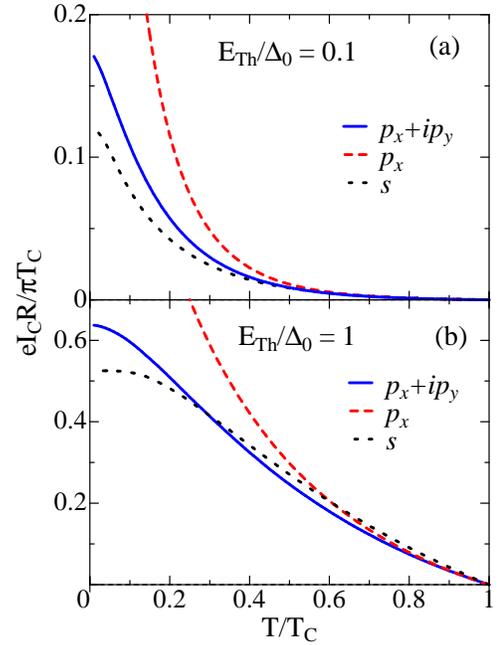}
  \end{center}
  \caption{(Color online) Temperature dependence of the critical current
for $Z=10$ and $R_d/R_b=1$. The solid lines are the results for
CP/DN/CP junctions, the broken lines are the results for P/DN/P
junctions, and the dot-lines are the results for S/DN/S junctions.
(a) $E_{\rm{Th}}/\Delta_{0}$= 0.1 and (b) $E_{\rm{Th}}/\Delta_{0}=1$.}
  \label{Comp1}
\end{figure}

In order to compare our results with the existing theories, we also
calculate $I_{\rm{C}}$ in S/DN/S junction and P/DN/P
junctions\cite{Yokoyama1,Yoko4}. The results are shown in Figure
\ref{Comp1}
for $Z=10$ and $R_d/R_b=1$. We find that the magnitude of $I_{\rm{C}}$
in CP/DN/CP junction is larger than that in S/DN/S junction, and
less than that in P/DN/P junction at low temperatures for both (a)
and (b). These results indicate that the maximum Josephson current
is enhanced due to the unusual proximity effect coexisting  with
MARS in CP/DN/CP junctions. However, it is known that MARS is
induced only for the particle with injection angle $\phi=0$ in
CP/DN/CP junctions, thus the $I_{\rm{C}}$ is smaller than in P/DN/P
junctions. We also find that qualitative temperature dependence of
the critical current in CP/DN/CP junctions is quite similar to that
in S/DN/S junctions. The result is consistent with the experiment in
Sr$_2$RuO$_4$-Sr$_3$Ru$_2$O$_7$ eutectic junctions\cite{Hooper}. As
follows from these calculations, if we focus on the temperature
dependence and current phase relation of the Josephson current of
CP/DN/CP junctions, the obtained results are rather conventional.

\subsection{CP / DF / CP junctions}
It is well known that in S/DF/S junctions, 0-$\pi$
transition\cite{Buzdin,Ryazanov} can be induced. This phenomenon is
due to the proximity effect specific to DF. In DF, the Cooper pairs
have finite center of mass momentum and the pair amplitude is
spatially oscillating. As a result, various interesting phenomena
are predicted in these
junctions\cite{buzdinrev,bverev,Yokoyama,Yoko,Yoko3}. The 0-$\pi$
transition is a typical example. It also exists in $d(p)$-wave
superconductor / DF / $d(p)$-wave superconductor
junctions\cite{Yoko4}. Here we study the Josephson effect in CP / DF
/ CP junctions. \par
\begin{figure}[htbp]
  \begin{center}
\includegraphics[keepaspectratio=true,height=90mm]{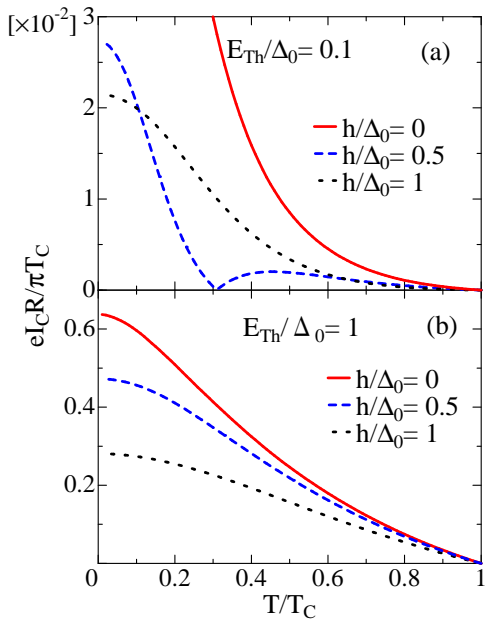}
  \end{center}
  \caption{(Color online) Temperature dependence of the critical current for $Z=10$ and $R_{d}/R_{b}=1$.
  The solid lines are the results for $h/\Delta_{0}=0$, the broken lines are the results for $h/\Delta_{0}=0.5$,
  and the dot-lines are the results for $h/\Delta_{0}=1$. (a) $E_{\rm{Th}}/\Delta_{0}=0.1$ and (b) $E_{\rm{Th}}/\Delta_{0}=1$.}
  \label{CCF1}
\end{figure}

 Figure \ref{CCF1} shows the temperature
dependence of the critical current for $Z=10$ and $R_{d}/R_{b}=1$. In
all cases, the exchange field suppresses the magnitude of $I_{\rm{C}}$.
For $E_{\rm{Th}}/\Delta_{0}=0.1$ and $h/\Delta_{0}=0.5$, the
non-monotonic temperature dependence of $I_{\rm{C}}$ is realized.

\begin{figure}[htbp]
  \begin{center}
\includegraphics[keepaspectratio=true,height=90mm]{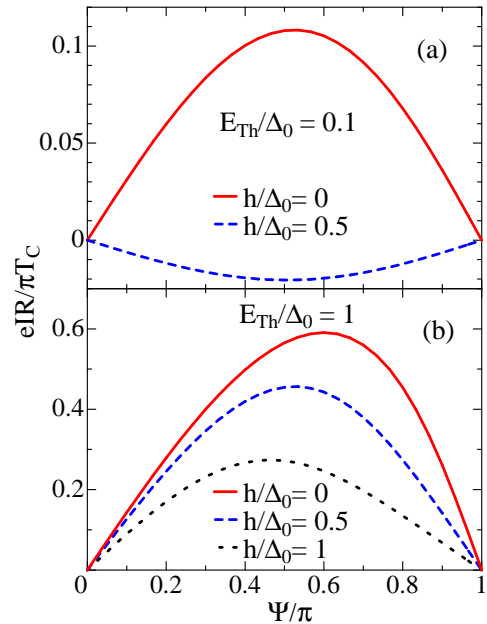}
  \end{center}
  \caption{(Color online) Current-phase relation for $Z=10$ and $R_d/R_b=1$ at $T/T_{\rm{C}}=0.1$. The solid lines are the results for $h/\Delta_{0}=0$,
  the broken lines are the results for $h/\Delta_{0}=0.5$, and the dot-line is the result for $h/\Delta_{0}=1$.
  (a) $E_{\rm{Th}}/\Delta_{0}=0.1$ and (b) $E_{\rm{Th}}/\Delta_{0}=1$.}
  \label{CPRF1}
\end{figure}

To clarify that this non-monotonic temperature dependence originates
from the 0-$\pi$ transition, we focus on the current phase relation
as shown in Figure \ref{CPRF1}  for $Z=10$ and $R_d/R_b=1$ at
$T/T_{\rm{C}}=0.1$. With the increase of the exchange field, the maximum
of the Josephson current is shifted to $\Psi<0.5\pi$ for
$E_{\rm{Th}}/\Delta_{0}=1$. Especially, for $E_{\rm{Th}}/\Delta_{0}=0.1$,
the Josephson current changes it sign for $h/\Delta_{0}=0.5$. These
results indicate that the exchange field induces the 0-$\pi$
transition in this case.

\begin{figure}[htbp]
  \begin{center}
\includegraphics[keepaspectratio=true,height=90mm]{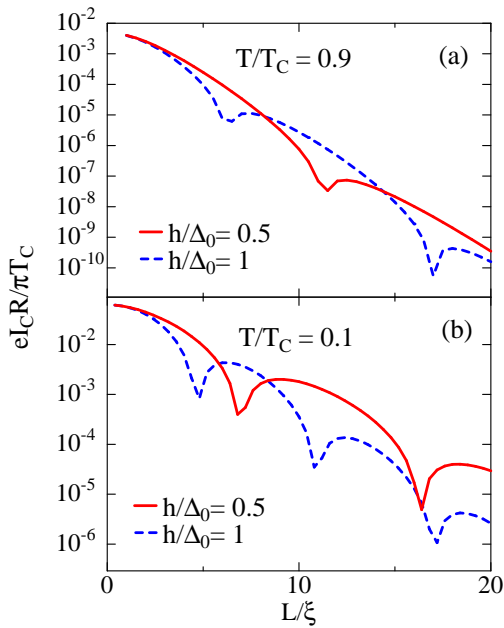}
  \end{center}
  \caption{(Color online) The critical current as a function of DF length for $Z=10$ and $R_d/R_b=1$. (a) $T/T_{\rm{C}}$= 0.9 and (b) $T/T_{\rm{C}}=0.1$.
  The solid lines are the results for $h/\Delta_{0}=0.5$ and the broken lines are the results for $h/\Delta_{0}=1$.}
  \label{Leng2}
\end{figure}

In Figure \ref{Leng2}, $I_{\rm{C}}$ is plotted  as a function of the
length $L$ of DF. In the presence of  the exchange field $h$ in DF,
the $I_{\rm{C}}$ oscillates as a function of length of DF. The period of
this oscillation becomes shorter with the increase of the magnitude
of $h$.

We have shown  that 0-$\pi$ transition also exists in the CP/DN/CP
junctions. The nonmonotonic temperature dependence of $I_{\rm{C}}$ and
the oscillatory dependence of $I_{\rm{C}}$ as a function of $L$ are
consistent with S/DN/S junctions or $d(p)$-wave superconductor / DF
/ $d(p)$-wave superconductor junctions\cite{Yoko4}. It is shown that
the  0-$\pi$ transition specific to DF junctions is robust against
the change of the symmetry of the Cooper pair.

\section{Conclusions}

We have derived the generalized boundary conditions for DN (DF) /
CP interface including the
macroscopic phase of the superconductor. The Josephson effect in CP
/ DN (DF) / CP junctions has been studied by solving the Usadel
equations with the above boundary conditions. Here, we choose the
$p_x+ip_y$-wave as the symmetry of CP superconductor. The results
obtained in the present paper can be summarized as follows:
\begin{enumerate}
\item It is shown that the symmetry of the
induced  pair wave function in DN due to the proximity effect is
odd-frequency spin triplet $s$-wave. Josephson current is carried by
the odd-frequency paring state.
\item
Almost all of the obtained results are qualitatively similar to
those in S/DN/S junctions. $I_{\rm{C}}$ is proportional to $\exp(-L/\xi)$
where $L$ and $\xi$ is the length of DN and coherence length of
Cooper pair in DN, respectively. The temperature dependence of the
maximum Josephson current in CP/DN/CP junction is qualitatively
similar to that in the S/DN/S junctions.
\item
Although the magnitude of the $I_{\rm{C}}$ is enhanced at low
temperatures as compared to the corresponding S/DN/S junctions, this
enhancement is not as strong as in the case of  P/DN/P junctions.
\item In CP/DF/CP junctions, current phase relation
changes drastically with the decrease of the temperature due to the
0-$\pi$ transition. The resulting $I_{\rm{C}}$ oscillates as a function
of the width of DF. These properties are similar to those of S/DN/S
junctions.
\end{enumerate}

Recently, the Josephson effect in Sr$_2$RuO$_4$-Sr$_3$Ru$_2$O$_7$
eutectic junction is experimentally observed\cite{Hooper}. There is
no qualitative difference of the temperature dependence as compared
to that of S/DN/S junctions. The present theoretical result is
consistent with this experiment. Surprisingly, although the
proximity effect is unusual due to the presence of odd-frequency
pairing state, the resulting Josephson current is not much different
compared to the conventional junctions. The reason is that in the
present case, the magnitude of the odd-frequency pair amplitude is
small compared to that in P/DN/P junctions. Especially, the
magnitude of the pair amplitude in DN for low Matsubara frequency in
the present CP/DN/CP junctions is much smaller than that of P/DN/P
junctions. It should be stressed that even though there is no
qualitative difference between the actual experimentally observed
Josephson current \cite{Hooper} and that in the S/DN/S junctions, it
means neither absence of the spin-triplet pairing state in
Sr$_{2}$RuO$_{4}$ nor absence of the odd-frequency pairing state in
DN.
\par
In the present paper, we only focus on the diffusive limit.
Recently, theory of proximity effect in the clean limit case is
presented \cite{Tanuma}. In such a case, the quasiclassical Green's
function should be described by Eilenberger equation. It is an
interesting issue to study the transition from clean limit to
diffusive limits systematically.

\section*{ACKNOWLEDGEMENT}

T. Y. acknowledges support by JSPS Research Fellowships for Young
Scientists. This work was supported by NAREGI Nanoscience Project,
the Ministry of Education, Culture, Sports, Science and Technology,
Japan, the Core Research for Evolutional Science and Technology
(CREST) of the Japan Science and Technology Corporation (JST) and a
Grant-in-Aid for the 21st Century COE "Frontiers of Computational
Science" . The computational aspect of this work has been performed
at the Research Center for Computational Science, Okazaki National
Research Institutes and the facilities of the Supercomputer Center,
Institute for Solid State Physics, University of Tokyo and the
Computer Center. This work is supported by Grant-in-Aid for
Scientific Research on Priority Area "Novel Quantum Phenomena
Specific to Anisotropic Superconductivity" (Grant No. 17071007) and
B (Grant No. 17340106) from the Ministry of Education, Culture,
Sports, Science and Technology of Japan.

%======Reference===================================

%===============================================================

\end{document}